\documentclass[letterpaper,12pt]{article}

\usepackage[top=1.0in,bottom=1.0in,left=1.0in,right=1.0in]{geometry}

\usepackage{amssymb}
\usepackage{amsmath}
\usepackage{graphicx}
\usepackage{url}
\usepackage{xspace}

\newcommand\lhood{{\cal L}}
\newcommand\mhiggs{m_\mathrm{H}}
\newcommand\mhiggshat{\hat m_\mathrm{H}}
\newcommand\ltheta{\ensuremath{\lhood(\theta)}\xspace}
\newcommand\lmass{\ensuremath{\lhood(\mhiggs)}\xspace}
\newcommand\lmasshat{\ensuremath{\lhood(\mhiggshat)}\xspace}

\newcommand\kgamma{\kappa_\gamma}
\newcommand\kgluon{\kappa_\mathrm{g}}
\newcommand\lkappa{\ensuremath{\lhood(\kgamma,\kgluon)}\xspace}

\newcommand\model{\ensuremath{p(x|\theta)}\xspace}

\begin{document}

\title{\bf \Large What is the likelihood function,\\ and how is it used in particle physics?}
\author{Robert D. Cousins\thanks{cousins@physics.ucla.edu}\\
Dept.\ of Physics and Astronomy\\ 
University of California, Los Angeles\\
Los Angeles, California 90095}
 
\date{October 1, 2020}

\maketitle

\begin{abstract}
Likelihood functions are ubiquitous in data analyses at the LHC and
elsewhere in particle physics.  Partly because ``probability'' and
``likelihood'' are virtual synonyms in everyday English, but crucially
distinct in data analysis, there is great potential for confusion.
Furthermore, each of various approaches to statistical inference
(likelihoodist, Neyman-Pearson, Bayesian) uses the likelihood function
in different ways. This note is intended to provide a brief
introduction at the advanced undergraduate or beginning graduate student
level, citing a few papers giving examples and containing
numerous pointers to the vast literature on likelihood. The Likelihood
Principle (routinely violated in particle physics analyses) is
mentioned as an unresolved issue in the philosophical foundations of
statistics.
\end{abstract}

\bigskip\noindent
Invited contribution to the newsletter of the CERN EP Department, September 2020,\\
\url{https://ep-news.web.cern.ch/node/3213} (reformatted in Latex with minor changes)

\clearpage
\section{Introduction}

In statistical data analysis, ``likelihood'' is a key concept that is
closely related to, but crucially distinct from, the concept of
``probability'', which is often a synonym in everyday English. In this
article, I first lay some necessary foundations for defining
likelihood, and then discuss some of its many uses in data analyses in
particle physics including the measured value of a parameter, an
interval expressing uncertainty, and hypothesis testing in various
forms. A simple example also illustrates how likelihood is central to
deep controversies in the philosophical foundations of statistical
inference. More complex examples at the LHC include measurements of
the properties of the Higgs boson.

\section{Statistical models}
Before delving into our discussion about the use of likelihood in
particle physics, we require the concept of ``statistical
models'', which are the mathematical expressions that relate
experimental observations to underlying theory or laws of physics. We
let $x$ stand for an arbitrarily large collection of observations such
as tracks of particles in the ATLAS or CMS detectors. We let $\theta$ stand
for a host of quantities (collectively called parameters) that include
constants of nature in putative or speculative laws of physics, as
well as all sorts of numbers that characterize the response of the
detectors. For example, calibration parameters relate flashes of light
in lead tungstate crystals to inferred amounts of energy deposited by
high-momentum electrons.

Even if all values of $\theta$ are kept fixed, the observations $x$ will
exhibit intrinsic variability, either because of slight changes in
conditions or more fundamentally because of the randomness in the laws
of quantum mechanics that govern the interactions in particle physics
and lead to statistical fluctuations. Therefore, statistical models,
usually written in the form of \model or $p(x;\theta)$, are not
deterministic equations but rather probabilities or probability
density functions (pdfs) for obtaining various $x$, given a set of
parameters $\theta$. Experiments provide observations of $x$, which are
used by scientists for inferences about components of $\theta$ (for
example the mass of the Higgs boson) or about the functional form of $p$
itself (such as whether the Higgs boson or other not-yet-seen
particles exist).

Defining ``probability'' is a long and difficult discussion beyond the
scope of this article. Here I simply assume that our readers most
likely learned the ``frequentist'' definition of probability (relative
frequency in an ensemble) in their quantum mechanics courses or
elsewhere, and use this definition in practice with Monte Carlo
simulation or efficiency measurements.

\section{The likelihood function}
After obtaining the collection of observations $x$, we take the
expression \model, evaluate it only for the specific $x$ that was
observed, and examine how it varies as $\theta$ is changed! This
defines the likelihood function \ltheta, often also denoted as
$\lhood(x|\theta)$ and sometimes $\lhood(\theta|x)$:
\begin{description}
\item{} \ltheta = \model, evaluated at the observed $x$, and considered to be a
function of $\theta$.
\end{description}

An additional overall arbitrary constant can be ignored.  Importantly,
while \model expresses probabilities or pdfs for various $x$, the
likelihood function \ltheta does {\em not} express probabilities or pdfs for
$\theta$. When Sir Ronald Fisher called it ``likelihood'' in the
1920s, his intent was to emphasize this important distinction. This
definition of the likelihood function is general enough to include
many special cases, for example where $x$ includes properties of
individual events (unbinned likelihood) as well where $x$ includes the
numbers of events in each bin of a histogram (binned likelihood).

Various uses of \ltheta are identified with at least three contrasting
approaches to statistical inference, which in the following are
referred to as ``likelihoodist'', ``Neyman-Pearson'' (NP), and
``Bayesian''. We illustrate them here with a statistical model \model
having a very simple mathematical form. But we must keep in mind that
when we analyze real experimental data, \ltheta is only as valid as
\model. Hence, great efforts go into specifying the mathematical form
of $p$ and obtaining values with uncertainties for the components of
$\theta$, which can number in the hundreds or even thousands. For this
purpose, massive ``calibration and alignment'' efforts for detectors
such as ATLAS and CMS are crucial, as are the many tests to ensure
that the models capture the essential aspects of the experimental
techniques. This need surely remains with the current trend toward
using ``machine learning'' methods, within which computer algorithms
develop numerical representations of \model that can be difficult for
humans to understand fully.

\section{The Likelihoodist approach}
For  illustration, we take as our statistical model the Poisson distribution that describes the probability of observing the integer $x$, which is the number of events that pass certain selection criteria. The value of $x$ is sampled from the Poisson distribution with an unknown true mean $\theta$:
\begin{equation}
\model = \theta^x e^{-\theta}/ x!.
\end{equation}

Now suppose that the value that was actually observed is $x$=3. This means that the likelihood function is

\begin{equation}
\ltheta = \theta^3 e^{-\theta}/ 6.
\end{equation}

The likelihoodist approach (advocated by A.W.F. Edwards in his 1972
monograph, {\em Likelihood}) takes the likelihood function as the
fundamental basis for the theory of inference. For example, the
likelihood ratio $\lhood(\theta_0)/\lhood(\theta_1)$ is an indicator
of whether the observation $x$=3 favors $\theta$=$\theta_0$ over
$\theta$=$\theta_1$.  Edwards considered values of $\theta$ for which
${-}2\ln\ltheta$ is less than one unit from the minimum value of
${-}2\ln\ltheta$ to be well supported by the observation. Figure~1
contains graphs of \ltheta and ${-}2\ln\ltheta$. For values of
$\theta$ at or near the maximum of \ltheta at $\theta$=3, the
observation $x$=3 had higher probability of occurring than for other
values of $\theta$.

\begin{figure} 
\begin{center}
\includegraphics[width=0.49\textwidth]{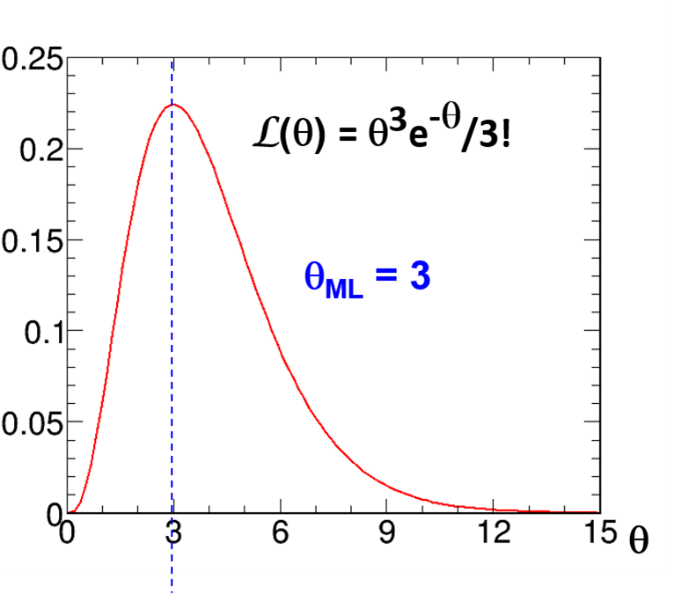}
\includegraphics[width=0.49\textwidth]{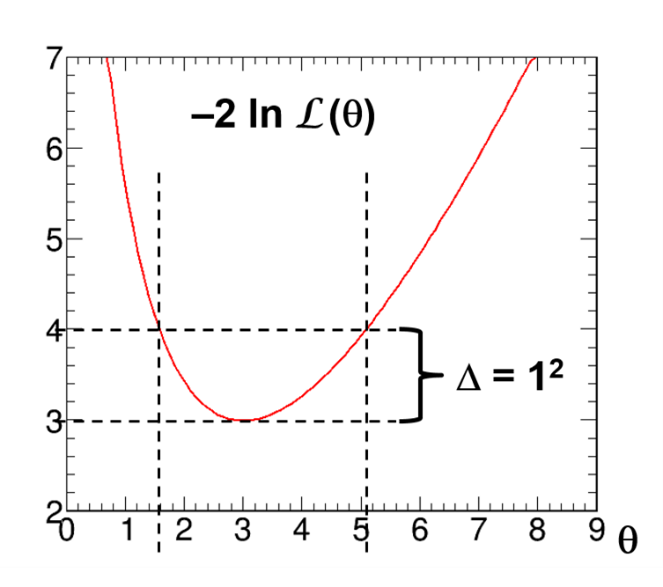}
\caption{
For the Poisson distribution, plots of the likelihood
function \ltheta and ${-}2\ln\ltheta$ in the case that $x$=3 is
observed. Since \ltheta is not a pdf in $\theta$, the area under \ltheta is
meaningless. The maximum likelihood estimate is
$\theta_\mathrm{ML}$=3, and an approximate 68\% CL confidence interval
is [1.58, 5.08], obtained from $\Delta({-}2\ln\ltheta)$=1.
}
\end{center}
\end{figure}  

\section{The Neyman-Pearson approach}
Most modern data analysts find that the pure likelihoodist approach is
too limiting, and that the likelihood (or ratio) needs to be
supplemented in order to have a calibrated interpretation as strength
of evidence.  In the 1920s and 30s, Neyman and Pearson (NP) developed
an approach to tests of hypotheses based on examining likelihood
ratios evaluated not only at the observed value of $x$, but also at
values of $x$ that could have been observed but were not. This was
generalized to constructing confidence intervals with associated
confidence levels (CL) that are common today.

We introduce the NP theory with a somewhat artificial example. Let's
assume that we have one physics hypothesis (H$_0$) that predicts
exactly $\theta$=$\theta_0$ and a second physics hypothesis ($H_1$)
that predicts exactly $\theta$=$\theta_1$. Neyman and Pearson
advocated choosing H$_0$ if the likelihood ratio
$\lhood(\theta_0)/\lhood(\theta_1)$ is above some cut-off value. This
cut-off value can be calculated from the experimenter's desired
maximum probability of choosing the second hypothesis if the first hypothesis
is true (the ``Type I error probability''). The Neyman-Pearson lemma says
that for any desired Type I error probability, choosing the correct
hypothesis based on a cutoff for this likelihood ratio will minimize the
probability that the first hypothesis is chosen when the second hypothesis is
true (the ``Type II error probability'').

A key aspect of this approach is that the calculations of the NP error
probabilities involve evaluating \model at all possible values of $x$
(the ``sample space''), not just at the observed value. In particle
physics, the ``number of $\sigma$'', or the equivalent $p$-value that
is used as a measure of strength of evidence for an observation or
discovery, is closely related, and shares this aspect.

In 1937, Neyman showed how to construct ``confidence intervals'' in
$\theta$ (his phrase). The lower and upper endpoints
$[\theta_\mathrm{L} , \theta_\mathrm{U}]$ are functions of the
observed $x$, as are Edwards's intervals of support. The defining new
feature was that one could choose a number called the ``confidence
level'' (CL); 68\% and 95\% are common at the LHC. With $x$ randomly
sampled from the sample space according to the statistical model, the
endpoints $[\theta_\mathrm{L} , \theta_\mathrm{U}]$ are distributed
such that the confidence intervals contain (``cover'') the unknown
true value of $\theta$ in at least a fraction CL of the intervals, in
the imagined limit of infinite repetitions. The phase ``at least'' is
necessary when the observations are integers; for continuous
observations, the coverage is exact in the simplest models.

In fact, the ensemble of confidence intervals need not be from the
same statistical model. In principle, the set of measurements in the
Particle Data Group's Review of Particle Physics could serve as the
ensemble, if all the statistical models were correctly
described. There are, however, enough incompatible measurements in the
listings to indicate that we have not achieved this ideal.

Meanwhile Fisher, who subscribed neither to the strict likelihoodist
philosophy nor to NP testing, proved multiple theorems about the
usefulness of using the ``method of maximum likelihood (ML)'' for
obtaining the quoted measured value of $\theta$ (``point
estimate''). He also studied the shape of the log-likelihood function
(second derivative, etc.)\ as measures of uncertainty, as well as
Fisher's definition of ``information'' contained in the observation.

A common point of view nowadays is that intervals of support based on
differences of ${-}2\ln\ltheta$, particularly as generalized by
S.S. Wilks and others to the cases with models having many parameters
$\theta$, can be extremely useful as algorithms for constructing
intervals and regions that approximately satisfy Neyman's coverage
criterion for confidence intervals.  This approximation generally gets
better as the data sample size increases.  For example, ATLAS and CMS,
separately and in combined results, relied heavily on log-likelihood
differences for measurements of the mass and couplings of Higgs bosons
to other particles~[1,2], as illustrated below.

\section{The Bayesian approach}
Finally, to introduce the Bayesian approach, we must consider a
definition of probability other than the frequentist definition. An
old view, which has attracted many adherents in the last half century,
is that probability is not something intrinsic to decaying particles
(or rolling dice), but rather resides in an individual's mind as
``subjective degree of belief''. Nowadays the moniker ``Bayesian''
typically refers to this definition of probability. Whatever one
thinks of this, it undoubtedly applies to more situations,
particularly in a case with no relevant ensemble, such as ``the
probability that the first person on Mars returns to Earth
alive''. One can define a continuous pdf for a presumed ``constant of
nature'' such as the mass of the Higgs boson, from which one can
calculates one's belief that the mass is within some specified
interval. (A frequentist pdf does not exist, or is useless.)

In the Bayesian framework, one has the ``prior pdf''
$p_\mathrm{prior}(\theta)$ that specifies relative belief in values of
$\theta$ before the observation $x$ is incorporated into belief.
Afterwards one has the ``posterior pdf'' $p_\mathrm{posterior}(\theta|x)$, given
the observed $x$. The celebrated Bayes's Theorem, which applies both
to frequentist probabilities (when they exist) and to Bayesian
probabilities, implies the proportionality,
\begin{equation}
p_\mathrm{posterior}(\theta|x)\ \propto\ \ltheta\ \times\ p_\mathrm{prior}(\theta).
\end{equation}
The likelihood function (which is not a pdf in $\theta$), relates the
before-and-after beliefs about $\theta$ in this simple way.  The
posterior pdf can then be used for a variety of purposes, including
``credible intervals'' that correspond to specified levels of belief.

Among the controversies with this approach, including whether Bayesian
probability is a useful concept for scientific communication, are
those involving the choice of the prior pdf. Choices representing an
individual's degree of belief may be objectionable to those with other
beliefs. A completely different way to obtain prior pdfs for $\theta$
is to use formal mathematical rules applied to the functional form of
the model \model~[3]. These so-called ``objective priors'' depend on
the measurement apparatus and protocol!

As an example, if a measurement apparatus has a Gaussian resolution
function for the mass, the usual formal rules dictate a prior pdf
uniform in mass. But if a different measurement apparatus has a
Gaussian resolution function for the mass-squared, then a prior pdf
uniform in mass-squared is dictated. By the rules of probability, the
latter implies a prior pdf for mass that is not uniform in mass. As
startling as that may seem, the formal rules can bring some perceived
advantages, and even help forge a connection to Neyman's frequentist
notion of coverage~[3].  It is a mistake, however, to think that it is
possible to fulfill an oft-expressed desire for a prior pdf that
expresses complete ignorance about $\theta$.

In particle physics, with its strong tradition of frequentist
coverage, prior pdfs are often chosen to provide intervals (in
particular upper limits for Poisson means) with good frequentist
coverage~[4]. In such cases, our use of Bayesian computational
machinery for interval estimation is not so much a change in paradigm
as it is a technical device for frequentist inference.

\section{The Likelihood Principle}
This simple Poisson model highlights a fundamental difference between,
on the one hand, NP testing and related concepts common in particle
physics, and on the other hand, the likelihoodist and Bayesian
approaches.  After data are obtained, the latter adhere to the
``Likelihood Principle'', which asserts that the model \model should
be evaluated only at the observed $x$.  In contrast, the usual
confidence intervals and measures of ``statistical significance''
reported in particle physics (3$\sigma$, 5$\sigma$, etc.) include
concepts such as the probability of obtaining data as extreme {\em or more
extreme} than that observed. This deep philosophical issue, which also
leads to differences in how one treats the criteria by which one stops
taking data, remains unresolved~[4].

In the limit of large samples of data, and if certain regularity
conditions are satisfied, the three approaches for measurements of
quantities such as the Higgs boson mass typically yield similar
numerical values, all having good coverage in the sense of Neyman, so
that any tension dissipates. In fact, both the likelihoodist and
Bayesian approaches can provide convenient recipes for obtaining
approximate confidence intervals.  However, the likelihoodist and
Bayesian conclusions when testing for a discovery can be quite
different from the usual ``number of $\sigma$'' in common use in
particle physics, even (especially!) with large data sets~[5].

\section{Beyond the simple Poisson example}
The discussion becomes substantially richer when there are two
observations of $x$, and/or two components of $\theta$, still within
the context of one statistical model~[4]. Another generalization is
when a second experiment has a different statistical model that
contains the same set of parameters $\theta$ and an independent set of
observations $y$. Then the likelihood function for the combined set of
observations $x$ and $y$ is simply the product of the likelihood
functions for each experiment. In the case of ATLAS and CMS, each
experiment has its own statistical model, reflecting differences in
the detectors, while having in common underlying laws of physics and
constants of nature.  A meta-model for the combination of the two
experiments requires understanding which parameters in each
statistical model are unique, and which correspond to parameters in
the other experiment's statistical model. For measurements of the
Higgs boson mass and properties, the collaborations invested
significant resources in preforming such combinations~[1,2].

Frequently, one is particularly interested in one or two components of
the set of parameters $\theta$, regardless of what the true values are
of all the other components of $\theta$.  Each of the three approaches
that presented here (likelihoodist, NP, Bayesian) has its associated
method(s) for ``eliminating'' these other ``nuisance parameters''
[4]. No method is entirely satisfactory, and sometimes a pragmatic
synthesis is preferred. The interested reader can pursue the topics of
``integrated likelihood'' and ``profile likelihood''. By convention,
the latter is commonly used by both ATLAS and CMS, including in the
figures shown here.

\begin{figure} 
\begin{center}
\includegraphics[width=0.6\textwidth]{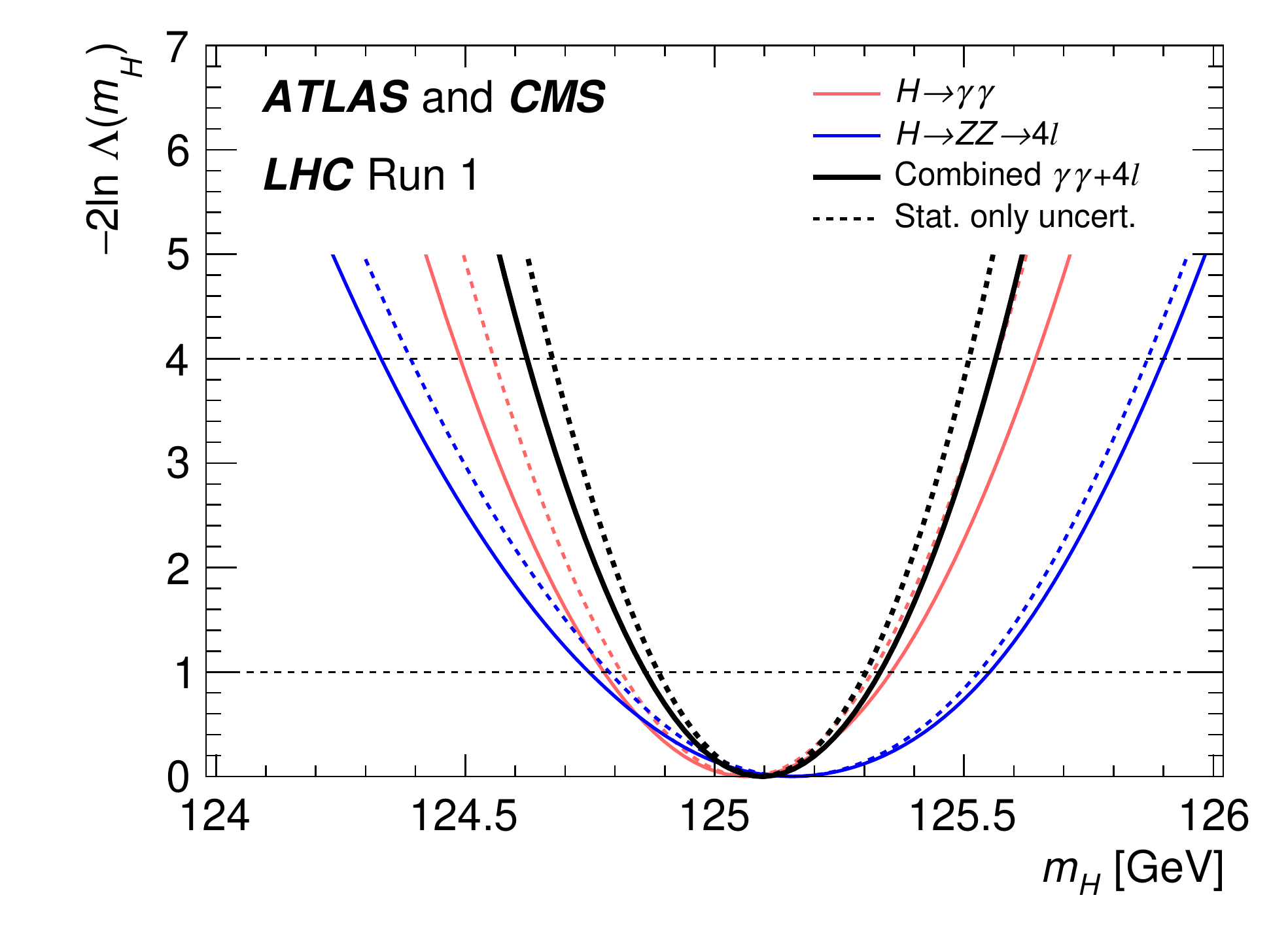} 
\caption{
Solid black curve: Measurement of the mass $\mhiggs$ of the
Higgs boson using the likelihood function \lmass from the first
combination of ATLAS and CMS data~[1]. The ML estimate is
$\mhiggshat$=125.09 GeV. The curves are ${-}2\ln\Lambda(\mhiggs)$,
where $\Lambda{=}\lmass/\lmasshat$, with nuisance parameters treated
by profile likelihood~[1] and suppressed in the notation
here. Approximate 68\% CL and 95\% CL confidence intervals are
indicated by the increases of 1 and 4 units. The other curves
indicated in the legend are described in Ref.~[1].
}
\end{center}
\end{figure}

\begin{figure} 
\begin{center}
\includegraphics[width=0.8\textwidth]{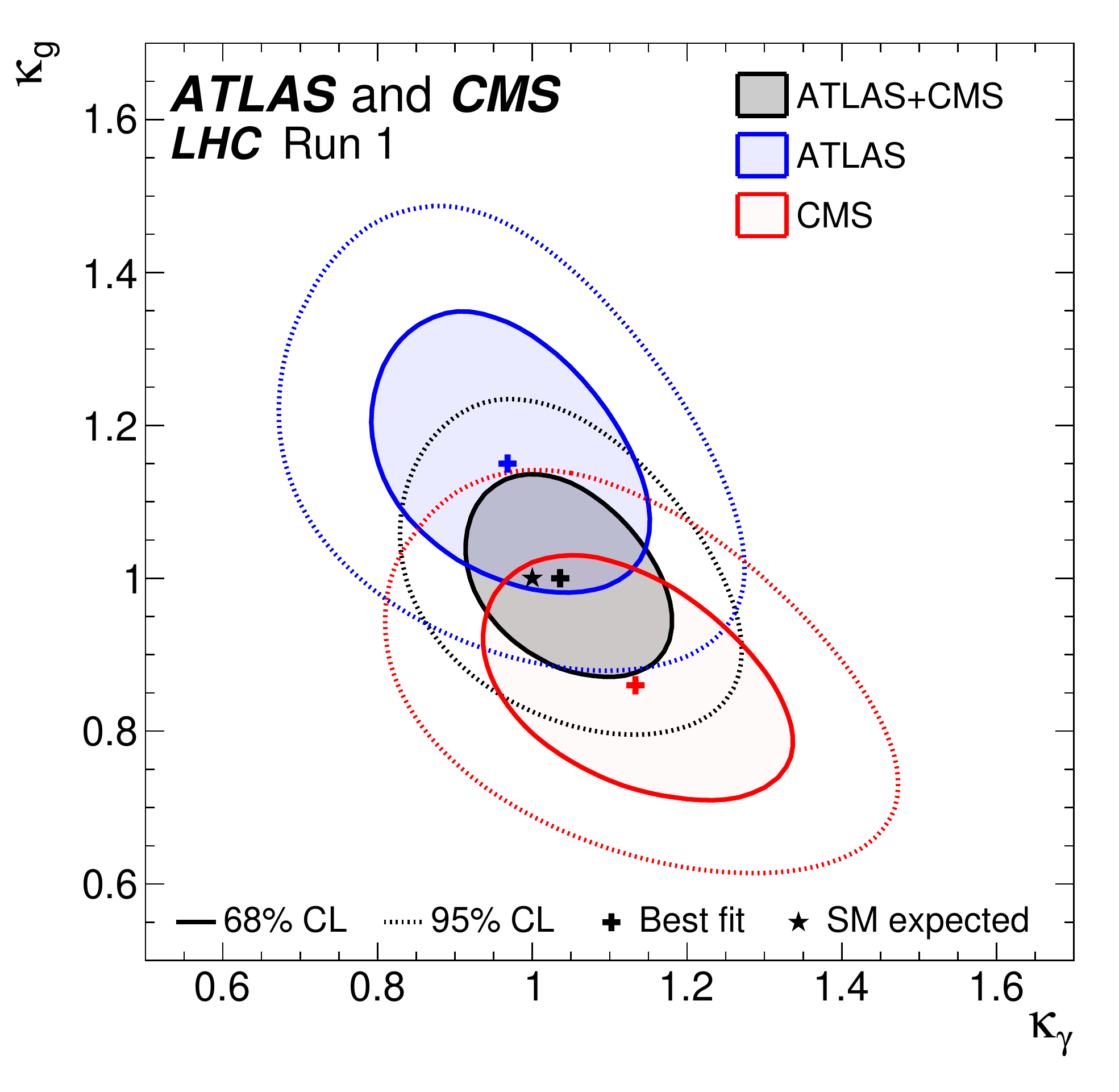} 
\caption{
For the two parameters of interest $(\kgamma,\kgluon)$ described in the
text, the curves in the plane are contours of ${-}2\ln\lkappa$ from the
combination of ATLAS and CMS data, and for each experiment separately~[2]. As described in Ref.~[2], the likelihood ratios used to define
each contour correspond to approximate 68\% CL and 95\% CL confidence
regions in the two-parameter plane. The crosses labeled ``Best fit''
indicate the ML estimates of the two parameters.
}
\end{center}
\end{figure}

Figures 2 and 3 are example plots for ATLAS-CMS combined measurements
from the first LHC run. In Fig.~2~[1], the parameter of interest was
the Higgs boson mass $\mhiggs$. Figure~3~[2] is part of a study of
great interest to see if the Standard Model (SM) correctly predicts
the effective coupling of Higgs bosons to two photons, as well as that
to two gluons. Two parameters of interest, $\kgamma$ and $\kgluon$,
were defined to be unity in the SM, and possible departures were
measured. With this initial level of precision, no significant
departure was observed.

In conclusion, we see that while the likelihood function is not a pdf
in the parameter(s) of interest, it has multiple uses in the various
approaches to statistical inference, all of which depend on an
adequate statistical model. There are many pointers to the likelihood
literature in the references below.

\section*{Acknowledgments}
I thank Tommaso Dorigo and Louis Lyons for comments on an early
version of the manuscript, and Panos Charitos for suggesting this
article and providing editorial support and suggestions.  This work
was partially supported by the U.S.\ Department of Energy under Award
Number {DE}--{SC}0009937.

\end{document}